\documentstyle[prl,aps,epsbox]{revtex}
\makeatletter
\begin{document}
\twocolumn

\newcommand{\Nc}{N_{\rm c}}

\def\slashchar#1{\setbox0=\hbox{$#1$}           
   \dimen0=\wd0                                 
   \setbox1=\hbox{/} \dimen1=\wd1               
   \ifdim\dimen0>\dimen1                        
      \rlap{\hbox to \dimen0{\hfil/\hfil}}      
      #1                                        
   \else                                        
      \rlap{\hbox to \dimen1{\hfil$#1$\hfil}}   
      /                                         
   \fi}                                         %

\title{Vector and chiral gauge theories on the lattice}
\author{Takanori Sugihara}%
\address{RIKEN BNL Research Center, 
Brookhaven National Laboratory, Upton, NY 11973, USA}
\maketitle

\begin{abstract}
We combine a pair of independent Weyl fermions 
to compose a Dirac fermion 
on the four-dimensional Euclidean lattice. 
The obtained Dirac operator is antihermitian and does not reproduce 
anomaly under the usual chiral transformation. 
To simulate the correct chiral anomaly, 
we modify the chiral transformation. 
We also show that chiral gauge theories can be constructed 
nonperturbatively with exact gauge invariance. 
The formulation is based on a doubler-free lattice derivative, 
which is a simple matrix defined as a discrete Fourier transform 
of momentum with antiperiodic boundary conditions. 
Long-range fermion hopping interactions are 
truncated using the Lanczos factor. 
\end{abstract}

\section{Introduction}

In the continuum Euclidean path-integral, chiral anomaly comes 
from Jacobian of the fermion measure \cite{fujikawa1}. 
The point is that ${\rm Tr}\gamma_5$ of the transformed 
fermion measure gives nonzero contribution 
because of the infinite degrees of freedom. 
On the other hand, lattice field theory is a framework 
to simulate the continuum theory 
with a finite number of lattice sites. 
Chiral anomaly cannot be reproduced on the finite lattice 
in the same way as the continuum theory. 
We should definitely distinguish finite and infinite 
lattice formulations and concentrate on reproducing 
chiral anomaly based on the finite lattice 
for practical numerical studies. 

The species doubling problem of the lattice fermion 
\cite{ks,slac,wilson,kaplan,Shamir:1993zy,neuberger,nn}
is closely related with chiral anomaly 
\cite{karsten,Karsten:1980wd,Seiler:1981jf,Ginsparg}. 
According to the Nielsen-Ninomiya theorem, 
a single Weyl fermion cannot exist on the lattice \cite{nn}. 
When formulating lattice Dirac fermion, 
one has almost no choice but to break chiral symmetry explicitly 
using Wilson terms \cite{wilson}. 
Otherwise, one has to give up one of the other 
assumptions of the theorem as seen in the literature.

L\"uscher's implementation of lattice chiral symmetry based on the 
Ginsparg-Wilson relation \cite{Ginsparg} was a breakthrough 
\cite{luscher}. 
The most significant feature to be stressed is that 
chiral anomaly can be devised even with a finite lattice. 
In the L\"uscher's formulation, 
new chiral transformation is introduced and 
chiral anomaly is obtained from Jacobian of the fermion measure 
in a different way from the continuum theory. 
The lattice index theorem holds for arbitrary lattice spacing 
\cite{Hasenfratz,luscher} and chiral anomaly agrees with 
the continuum result in the continuum limit 
\cite{kikukawa,fujikawa2,suzuki,adams}. 
The lesson to be learned is that one can modify 
the axial current in order to reproduce the correct 
chiral anomaly on the finite lattice. 

In the electroweak theory, left- and right-handed fermions 
couple to gauge fields in different ways. This type of 
formulation is called chiral gauge theory. 
For consistent construction of chiral gauge theories, 
gauge symmetry needs to be maintained at the quantum level 
(see Ref. \cite{bertlmann}, for example). 
Introduction of Wilson terms has trouble 
because the mixing of left- and right-handed 
fermions complicates discussion of 
gauge anomaly cancelation \cite{gaugeanomaly,Luscher:2000hn}. 
SLAC fermion partially simplifies the problem 
because it does not use Wilson terms \cite{slac,Melnikov:2000cc}. 
However, it has not been successful due to breakdown of 
locality and  Lorentz invariance associated with 
axial currents in the continuum limit \cite{karsten}. 
There have been discussions that deny these defects 
and the non-conservation of the axial current 
\cite{Rabin:1981nm,Ninomiya:hd}. 
The defects originate in the definition of axial currents 
with a derivative different from the SLAC derivative. 
Careful and consistent treatment is necessary when discussing 
species doubler and chiral anomaly with SLAC fermion. 

In this paper, we give a method to save the SLAC derivative. 
On a finite lattice, a derivative is defined as a discrete Fourier 
transform of momentum in a similar way to 
the conventional SLAC derivative. 
To remove doubler modes, antiperiodic boundary conditions are chosen 
for the derivative in real space. 
The doubler modes at the momentum boundary are lifted and the 
dispersion relation of the continuum theory is reproduced discretely. 
Although long-range hopping interactions appear, 
the pathology of SLAC fermion is avoided 
and the correct continuum limit is guaranteed. 
However, long-range interactions are not useful in practical 
numerical calculations because sparser Dirac operator 
is better for numerical efficiency. 
By using the Lanczos factor technique demonstrated 
in Ref. \cite{sugihara}, we effectively truncate long-range hopping 
interactions and improve the fermion propagator. 
Since the proposed lattice derivative does not use Wilson terms, 
left- and right-handed fermions are completely independent. 
As a result, the Dirac operator constructed with the derivative 
has exact chiral symmetry. 
This means that the Dirac fermion does not reproduce chiral 
anomaly because the fermion measure of path-integral is trivially 
invariant under the chiral transformation on the finite lattice. 
To simulate the correct chiral anomaly, 
we modify chiral transformation 
using the Neuberger's solution to the Ginsparg-Wilson relation 
\cite{neuberger}. 
As a result, we obtain an anomalous Ward identity for the modified 
chiral transformation. The zero mode of the axial current divergence 
is generally non-zero and gives the index theorem. 
The absence of chiral anomaly under the usual chiral transformation 
implies the existence of a single anomaly-free Weyl fermion. 
Chiral gauge theories can be constructed 
using the single Weyl fermion as a building block.

This work is a tricky extension of SLAC fermion. 
In the conventional SLAC fermion, 
periodic boundary conditions and an infinite lattice are assumed, 
where the doubler remains as a singularity at the momentum boundary. 
On the other hand, our formulation 
is based on antiperiodic boundary conditions 
and a finite lattice and therefore free from species doubler. 
In addition, the axial current is defined 
with the modified chiral transformation and 
evaluated nonperturbatively. 
The Nielsen-Ninomiya theorem does not apply to our implementation 
of chiral anomaly because the modified chiral transformation is 
used to define a symmetry.

This paper is organized as follows. 
In Sec. \ref{lattice_derivative}, the lattice derivative is defined 
based on the finite lattice formulation. 
Locality of the derivative is discussed. 
Long-range hopping interactions are truncated 
using the Lanczos factor. 
In Sec. \ref{chiral_anomaly}, a modified chiral transformation is 
introduced to reproduce the correct chiral anomaly. 
A way of constructing gauge-invariant chiral gauge theories 
is given in a nonperturbative way. 
Sec. \ref{sumamry_and_discussions} is devoted to 
summary and discussions. 

\section{Lattice derivative}

\label{lattice_derivative}
We define lattice derivative 
as a discrete Fourier transform of momentum 
\begin{equation}
 \nabla_n = \frac{1}{N} \sum_{l=-N/2+1}^{N/2} ip_l e^{i2\pi \tilde{l}n/N}
 = -\frac{2}{N} \sum_{l=1}^{N/2} p_l 
  \sin \left(\frac{2\pi\tilde{l}n}{N}\right), 
\label{nabla}
\end{equation}
where $n$ represents lattice sites and 
takes integer values between $-N/2+1$ and $N/2$. 
The lattice size $N$ is a finite even number and 
$p_l$ corresponds to momentum. 
\[
 p_l\equiv \frac{2\pi \tilde{l}}{N},\quad
 \tilde{l}\equiv l-\frac{1}{2}. 
\]
Antiperiodic boundary conditions have been chosen in real space, 
$\nabla_{n+N}=-\nabla_n$. 
\footnote{If periodic boundary conditions are chosen, 
there appear degenerate zero modes on the momentum boundary. 
Such unphysical doubler modes must be removed 
because they cause errors especially when fermion mass is small.}
In Eq. (\ref{nabla}), the summation can be carried out easily. 
\begin{equation}
 \nabla_n
 = \frac{\pi}{N^2}
 \left[
 (N+1)
 \frac{\cos\left(\displaystyle\frac{Ns}{2}\right)}
      {\sin\left(\displaystyle\frac{s}{2}\right)}
  -\frac{\sin\left(\displaystyle\frac{(N+1)s}{2}\right)}
       {\sin^2\left(\displaystyle\frac{s}{2}\right)}
 \right],
\end{equation}
where $s=2\pi n/N$. The lattice derivative 
(\ref{nabla}) is doubler-free and reproduces 
discretely the dispersion relation of the continuum theory 
for arbitrary lattice spacing. 

In the large $N$ limit, 
Eq. (\ref{nabla}) becomes an integral
\begin{equation}
 \frac{1}{a}\nabla_n = \frac{a}{2\pi} \frac{\partial}{\partial x}
  \int_{-\pi/a}^{\pi/a} dk \; e^{ikx}, 
 \label{nabla2}
\end{equation}
where $a$ is a lattice spacing and $x=na$ is a space coordinate. 
In the continuum limit $a\to 0$, we obtain the first order 
derivative of the continuum theory.
\begin{equation}
 \lim_{a\to 0}\frac{1}{a}\nabla_n 
 = a \frac{\partial}{\partial x} \delta(x). 
 \label{nabla3}
\end{equation}
The lattice derivative (\ref{nabla}) is local in the continuum limit.

\begin{figure}
  \begin{center}
    \epsfile{file=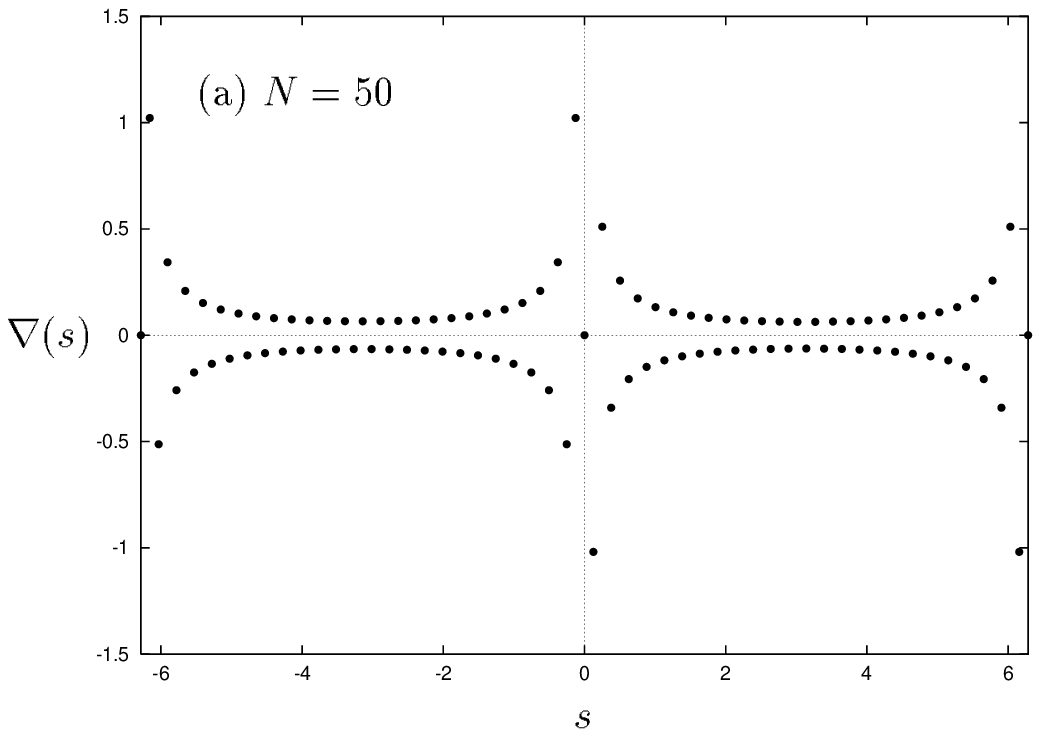,scale=0.8}
    \epsfile{file=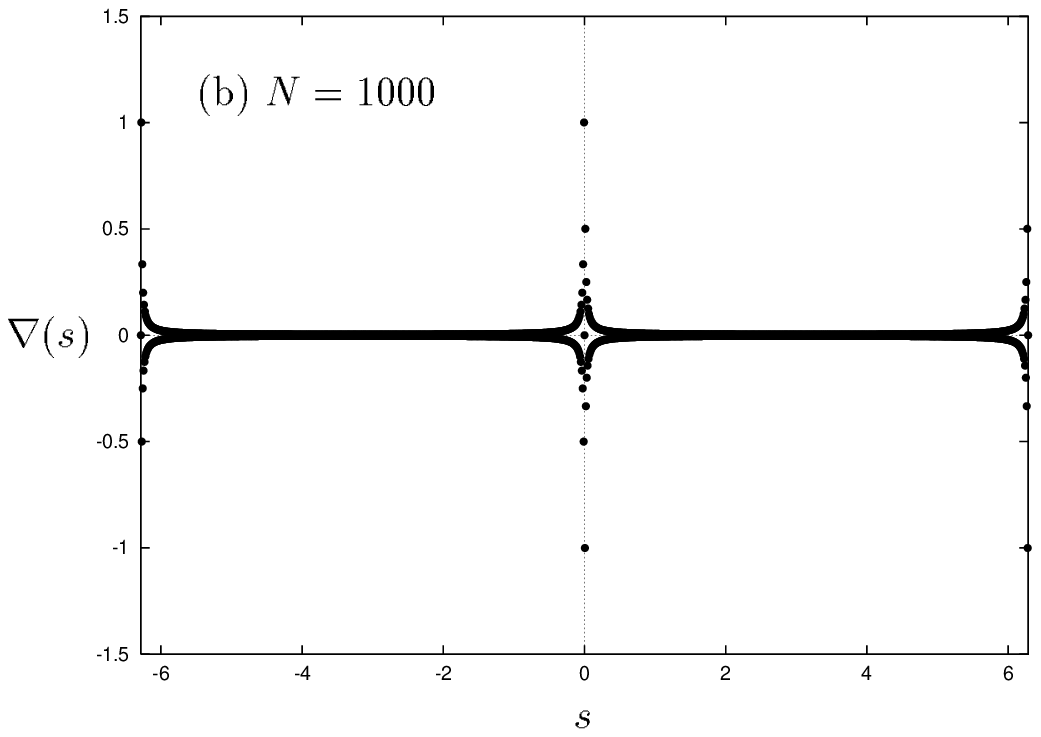,scale=0.8}
  \end{center}
\caption{
The lattice derivative $\nabla(s)\equiv\nabla_n$ is plotted as 
a function of $s=2\pi n/N$ for (a) $N=50$ and (b) $N=1000$ 
in a region $|s|\le 2\pi$. 
The function $\nabla(s)$ has antiperiodicity of $2\pi$, 
$\nabla(s+2\pi)=-\nabla(s)$.}
\label{fig1}
\end{figure}

Figure \ref{fig1} plots the lattice derivative 
$\nabla(s)\equiv\nabla_n$ with $s=2\pi n/N$ 
in a range $|s|\le 2\pi$ 
for two lattice sizes $N=50$ and $1000$. 
The absolute value of $\nabla(s)$ is large around 
$|s|=0$ and $2\pi$ and small around $|s|=\pi$. 
The points $|s|=0$ and $2\pi$ are equivalent because of 
antiperiodicity. 
Therefore, $\nabla(s)$ around $|s|=2\pi$ does not mean 
severe nonlocality. 
The points $|s|=\pi$ give the most long-range hopping. 
As expected from Eq. (\ref{nabla3}), 
locality of the derivative $\nabla(s)$ is quite good when $N=1000$. 
On the other hand, decay of the derivative $\nabla(s)$ is slow 
when the lattice size is small.

\begin{figure}
  \begin{center}
    \epsfile{file=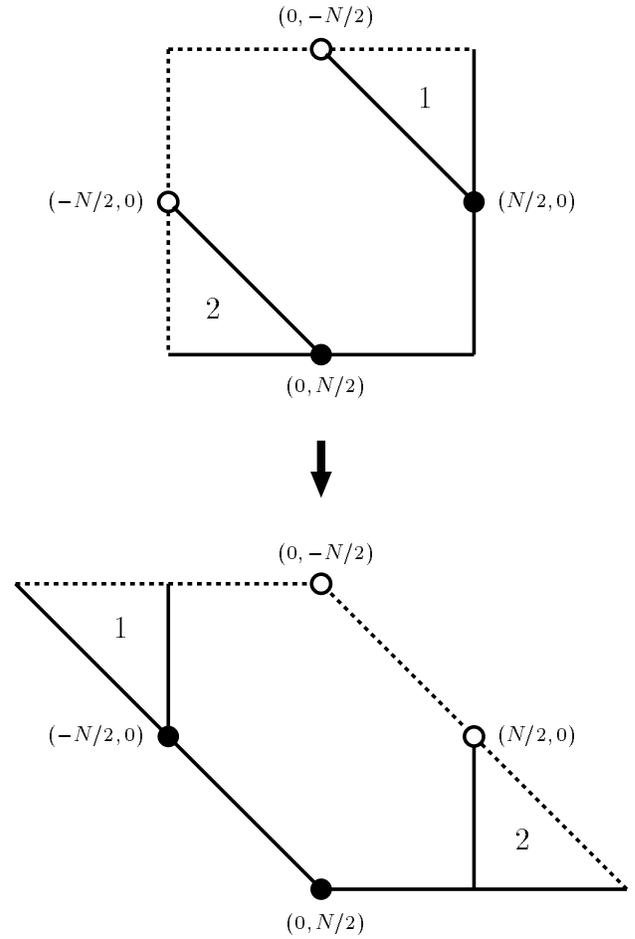,scale=1.0}
  \end{center}
\caption{The matrix $C_{m,n}$ is schematically shown 
(the upper diagram). 
The numbers in brackets are examples of the indices $(m,n)$. 
The matrix elements on the dotted lines are not contained 
in Eq. (\ref{pll1}). The two triangles 1 and 2 can be moved 
to form the parallelogram shape (the lower diagram). 
}
\label{fig2}
\end{figure}

We are interested in constructing a theory with better locality 
for a practical purpose. 
In order for Eq. (\ref{nabla}) to be a useful lattice derivative, 
long-range hopping interactions need to be truncated. 
However, truncation of hopping interactions may cause errors. 
We need to find a systematic way to reproduce 
spectra effectively with only short-range hopping interactions. 
For simplicity, let us consider a classical action 
for a free massless fermion in one-dimensional space. 
\begin{equation}
 S=a\sum_{m,n=-N/2+1}^{N/2}
 \bar{\psi}_m \frac{1}{a}\nabla_{m-n} \psi_n. 
\end{equation}
Fourier transforms of the one-component fermions are 
\begin{eqnarray*}
 \displaystyle
   \psi_n &=& \frac{1}{\sqrt{N}} \sum_{l=-N/2+1}^{N/2}
   e^{i2\pi\tilde{l}n/N} \zeta_l,
\\
 \displaystyle
   \bar{\psi}_n &=& \frac{1}{\sqrt{N}} \sum_{l=-N/2+1}^{N/2}
   e^{-i2\pi\tilde{l}n/N} \bar{\zeta}_l,
\end{eqnarray*}
where antiperiodicity is assumed
\[
 \psi_{n+N}=-\psi_n,\quad
 \bar{\psi}_{n+N}=-\bar{\psi}_n. 
\]
Then we have
\begin{equation}
 S=\sum_{l,l'} \bar{\zeta}_l ip_{l,l'} \zeta_{l'}, 
\end{equation}
where
\begin{eqnarray}
 p_{l,l'} &=& -\frac{i}{N}\sum_{m,n=-N/2+1}^{N/2}
  \nabla_{m-n} e^{-i2\pi\tilde{l}m/N+i2\pi\tilde{l}'n/N}
\nonumber
\\
 &=& p_l \delta_{l,l'}. 
\label{pll1}
\end{eqnarray}
with the inverse transform $p_l$ of Eq. (\ref{nabla}) 
\begin{equation}
 p_l = -2\sum_{n=1}^{N/2-1} \nabla_n
  \sin\left(\frac{2\pi\tilde{l}n}{N}\right)+(-1)^l\nabla_{N/2}. 
\label{pl0}
\end{equation}
To obtain this, antiperiodicity of $\nabla_n$, 
$\psi_n$, and $\bar{\psi}_n$ has been used. 
Antiperiodicity in real space gives rise to 
periodicity in momentum space, 
$p_{l+N}=p_l$, $\zeta_{l+N}=\zeta_l$, 
and $\bar{\zeta}_{l+N}=\bar{\zeta}_l$. 
Some explanations would be necessary 
for the derivation of Eq. (\ref{pl0}). 
Consider the matrix contained in Eq. (\ref{pll1}) 
\[
 C_{m,n}\equiv
 \nabla_{m-n} e^{-i2\pi\tilde{l}m/N+i2\pi\tilde{l}'n/N}, 
\]
which has periodicity, $C_{m+N,n}=C_{m,n+N}=C_{m,n}$. 
In Fig. \ref{fig2}, the matrix $C_{m,n}$ is shown schematically 
(see the upper diagram). 
Some examples for the indices $(m,n)$ are given. 
The matrix elements on the dotted lines 
are not contained in Eq. (\ref{pll1}). 
In the upper diagram, 
the vertices $(N/2,-N/2)$ and $(-N/2,N/2)$ 
correspond to the points with $|s|=2\pi$ in Fig. \ref{fig2}. 
Using the periodicity of the matrix, 
the triangles 1 and 2 can be moved to form the parallelogram 
(see the lower diagram). 
As a result, Eq. (\ref{pll1}) can be evaluated 
by calculating the summation for each $|m-n|$, 
which draws a line segment parallel to the oblique sides 
of the parallelogram. 
The last term of Eq. (\ref{pl0}) is a contribution of 
the term with $|m-n|=N/2$, which corresponds to 
the left oblique side of the parallelogram.

In Eq. (\ref{pl0}), we truncate long-range terms 
with a parameter $N_{\rm c}\le N/2$, 
which represents the largest distance of fermion hopping. 
\begin{equation}
 p_l = \sum_{n=1}^{N_{\rm c}} \nabla_n
  \Bigg[ (\delta_{n,\frac{N}{2}}-1)2
  \sin\left(\frac{2\pi\tilde{l}n}{N}\right)
  +\delta_{n,\frac{N}{2}}(-1)^l\Bigg]. 
\label{pl1}
\end{equation}
The truncation can be implemented to Eq. (\ref{pll1}). 
\begin{equation}
 p_{l,l'} = -\frac{i}{N} \sum'_{m,n}
  \nabla_{m-n} e^{-i2\pi\tilde{l}m/N+i2\pi\tilde{l}'n/N}. 
\label{pll2}
\end{equation}
The indices $m$ and $n$ run from $-N/2+1$ to $N/2$. 
The prime symbol means that the summation is taken over 
$|m-n|\le N_{\rm c}$, $m-n+N\le N_{\rm c}$, and $m-n-N\ge -N_{\rm c}$. 
Fermion hopping has been restricted to a finite range.

\begin{figure}
  \begin{center}
    \epsfile{file=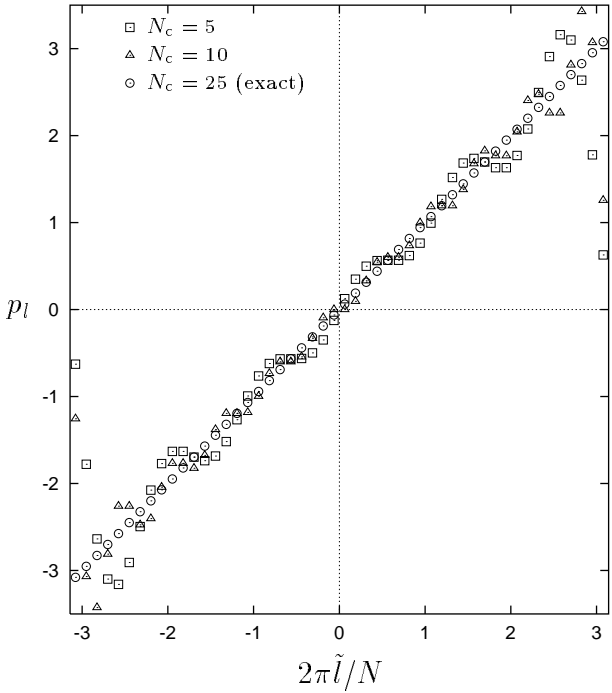,scale=1.0}
  \end{center}
\caption{Momentum $p_l$ of Eq. (\ref{pl1}) is 
plotted as a function of $2\pi\tilde{l}/N$ 
for $N_{\rm c}=5$ (squares), $10$ (triangles), and $25$ (circles) 
with a lattice size $N=50$. $N_{\rm c}=25$ gives the exact result. 
The truncated results oscillate around the exact one. 
}
\label{fig3}
\end{figure}

Figure \ref{fig3} plots $p_l$ of Eq. (\ref{pl1}) 
as a function of $2\pi\tilde{l}/N$ 
for $N_{\rm c}=5$, $10$, and $25$ with a lattice size $N=50$. 
$N_{\rm c}=25$ gives the exact result with no truncation, 
which satisfies the dispersion relation of the continuum theory. 
If one does not mind inclusion of long-range hopping, 
doubler-free formulation of a single Weyl fermion 
is possible maintaining the correct dispersion relation. 
When $N_{\rm c}$ is small, there is also no doubler modes 
because of antiperiodic boundary conditions. 
Although some modes around the momentum boundary 
deviate from the correct dispersion, 
those are not so harmful 
because there is no genuine degeneracy with low-lying modes. 
However, $p_l$ oscillates around the exact result. 
The oscillation tends to be large as $N_{\rm c}$ goes to small.

The small oscillation around the correct dispersion comes from 
the truncated terms having large $n$'s in Eq. (\ref{pl1}). 
As shown in Ref.~\cite{sugihara}, such oscillation can be 
removed by introducing the Lanczos factor, which is used 
in Fourier analysis to cancel the Gibbs phenomenon. \cite{aw} 
We modify Eq. (\ref{pl1}) as follows: 
\begin{equation}
 p_l = \sum_{n=1}^{N_{\rm c}} F_n \nabla_n
  \Bigg[ (\delta_{n,\frac{N}{2}}-1)2
  \sin\left(\frac{2\pi\tilde{l}n}{N}\right)
  +\delta_{n,\frac{N}{2}}(-1)^l\Bigg]. 
\label{pl2}
\end{equation}
where
\begin{equation}
 F_n \equiv \frac{N_{\rm c} +1}{\pi n}
  \sin\left(\frac{\pi n}{N_{\rm c} +1}\right). 
\end{equation}
is the Lanczos factor. As a result, Eq. (\ref{pll2}) becomes 
\begin{equation}
 p_{l,l'} = -\frac{i}{N} \sum'_{m,n} F_{|m-n|}
  \nabla_{m-n} e^{-i2\pi\tilde{l}m/N+i2\pi\tilde{l}'n/N}. 
\label{pll3}
\end{equation}
The final form of the action is given as
\begin{equation}
 S=a\sum'_{m,n}
 \bar{\psi}_m \frac{1}{a}F_{|m-n|}\nabla_{m-n} \psi_n. 
\label{action1}
\end{equation}

\begin{figure}
  \begin{center}
    \epsfile{file=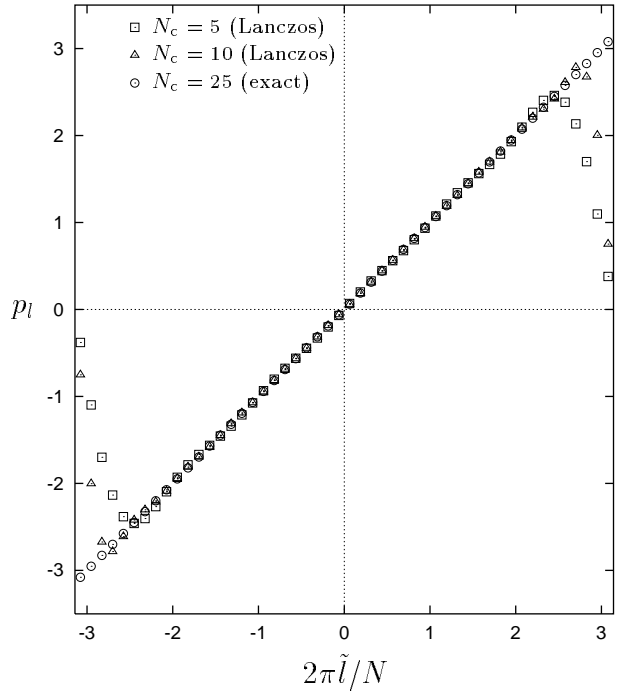,scale=1.0}
  \end{center}
\caption{Momentum $p_l$ of Eq. (\ref{pl2}) 
is plotted as a function of $2\pi\tilde{l}/N$ 
for $N_{\rm c}=5$ (squares) and $10$ (triangles) 
with a lattice size $N=50$. 
The result for $N_{\rm c}=25$ (circles) is the exact one 
shown in Fig. \ref{fig3}. 
The oscillation around the exact result has been removed 
with the Lanczos factor. 
}
\label{fig4}
\end{figure}

Figure \ref{fig4} plots $p_l$ of Eq. (\ref{pl2}) improved with 
the Lanczos factor as a function of $2\pi\tilde{l}/N$ 
for $N_{\rm c}=5$, $10$, and $25$ with $N=50$, 
which are compared with the exact result ($N_{\rm c}=25$) 
shown in Fig. \ref{fig3}. As before, 
there is no doubler for every $N_{\rm c}$. 
In addition to this, 
the oscillation has been removed with the Lanczos factor. 
As $N_{\rm c}$ goes to large, the deviation 
around the momentum boundary tends to be small. 
In this way, we can construct a doubler-free ultralocal derivative. 
If the Lanczos factor is introduced, 
ultralocal formulation of a single Weyl fermion is possible 
maintaining almost correct dispersion relation. 

\section{Chiral and gauge anomaly}

\label{chiral_anomaly}
On the finite four-dimensional Euclidean lattice, 
consider an effective action $\Gamma[U]$ 
\begin{equation}
 e^{-\Gamma[U]}=\int {\cal D}\psi {\cal D}\bar{\psi} e^{-S[U]}, 
\label{ea}
\end{equation}
where 
\begin{equation}
 S=a^4\sum_{m,n} \bar{\psi}_m \slashchar{D}_{m,n} \psi_n, 
\end{equation}
is a classical action for a massless Dirac fermion 
coupled to gauge. 
$\slashchar{D}\equiv \gamma_\mu D_\mu$ is 
a Dirac operator and the Euclidean Dirac matrices 
satisfy $\gamma_\mu^\dagger=\gamma_\mu$ and 
$\{\gamma_\mu,\gamma_\nu\}=2 \delta_{\mu\nu}$. 
Chirality is defined with $\gamma_5\equiv \gamma_1\gamma_2\gamma_3\gamma_4$. 
The fermion variables $\bar{\psi}_m$ and $\psi_n$ 
are Grassmann valued. The indices $m$ and $n$ are 
four-component numbers to indicate lattice sites and 
each component runs from $-N/2+1$ to $N/2$ as before. 
The lattice covariant derivative
\begin{equation}
 (D_\mu)_{m,n} \equiv \frac{1}{a}
  \nabla_{m_\mu-n_\mu} U_{m,n}(\mu)
  \prod_{\nu=1 (\nu\ne\mu)}^4 \delta_{m_\nu,n_\nu}
\label{lcd}
\end{equation}
is diagonal with respect to the space-time indices $m$ and $n$ 
except for the $\mu$-th ones. 
The Dirac operator $\slashchar{D}$ is antihermitian and 
satisfies $\{\slashchar{D},\gamma_5\}=0$. 
The classical action is invariant 
under the usual chiral transformation. 
The derivative $\nabla_n$ can be replaced with 
the truncated one in the same way as Eq. (\ref{action1}), 
if ultralocal construction is preferred. 
The gauge variable $U_{m,n}(\mu)$ is a product of all link variables 
that compose a line segment between the two sites $m$ and $n$ 
parallel to the $\mu$-th direction. 
When connecting the two sites with link variables 
along the $\mu$-th direction, there are two ways
because of periodicity of the action. 
The most natural choice is the shorter path. 
One of the two ways is chosen depending on the distance 
between two sites. 
When $|m_\mu-n_\mu|\le N/2$, 
\begin{equation}
 U_{m,n}(\mu)=\left\{
  \begin{array}{ll}
  U_{m,m-\hat{\mu}}\dots U_{n+\hat{\mu},n} &(m_\mu>n_\mu)\\
  U_{m,m+\hat{\mu}}\dots U_{n-\hat{\mu},n} &(m_\mu<n_\mu)\\
  \end{array}\right.,
\end{equation}
which corresponds to the hexagon 
sandwiched between the triangles 1 and 2 in Fig. \ref{fig2}. 
$\hat{\mu}$ is a unit vector in the $\mu$-th direction. 
When $|m_\mu-n_\mu|>N/2$, 
\begin{equation}
 U_{m,n}(\mu)=\left\{
  \begin{array}{ll}
  U_{m,m+\hat{\mu}}\dots U_{n+(N-1)\hat{\mu},n+N\hat{\mu}}
  &(m_\mu>n_\mu)\\
  U_{m,m-\hat{\mu}}\dots U_{n-(N+1)\hat{\mu},n-N\hat{\mu}}
  &(m_\mu<n_\mu)\\
  \end{array}\right.,
\end{equation}
which corresponds to the triangles 1 and 2 in Fig. \ref{fig2} 
and therefore intersects the boundary. 
The link variables $U_{n+\hat{\mu},n}$ are elements of a gauge group 
and satisfy  $U_{n,n+\hat{\mu}}=U_{n+\hat{\mu},n}^\dagger$ and 
$U_{n+\hat{\mu}+N\hat{\mu},n+N\hat{\mu}}=U_{n+\hat{\mu},n}$. 
In the continuum limit, 
the lattice covariant derivative becomes
\begin{equation}
 \lim_{a\to 0} (D_\mu)_{m,n}=a^4\delta^4(x-y)D_\mu^{({\rm c})}, 
\label{cntlcd}
\end{equation}
where $x=ma$ and $y=na$,
and $D_\mu^{({\rm c})}=\partial_\mu-igA_\mu$ 
with a parameterization $U_{m+\hat{\mu},m}=e^{iag A_\mu(x)}$. 
Eq. (\ref{lcd}) reduces to the covariant derivative 
of the continuum theory in the continuum limit. 

With our Dirac operator, the usual chiral transformation 
does not reproduce chiral anomaly because the classical action 
and the fermion measure is invariant. 
To simulate the correct chiral anomaly on the finite lattice, 
we introduce the modified chiral transformation with $\theta_n\ll 1$ 
\begin{eqnarray}
 \psi'_m &=&
 \sum_n \left(1+i\theta_m\hat{\gamma}_5\right)_{m,n}\psi_n,
\label{nct1}
\\
 \bar{\psi}'_m &=&
 \sum_n \bar{\psi}_n\left(1+i\theta_m\hat{\gamma}_5\right)_{n,m},
\label{nct2}
\end{eqnarray}
where 
\begin{equation}
 (\hat{\gamma}_5)_{m,n} \equiv \gamma_5
 \left(1-\frac{1}{2}a G \right)_{m,n}. 
\end{equation}
The operator $G$ is the Neuberger's solution \cite{neuberger} to the 
Ginsparg-Wilson relation $\gamma_5 G + G \gamma_5 = aG \gamma_5 G$ 
and has nothing to do with the Dirac operator (\ref{lcd}). 
The modified chiral transformation is a symmetry of 
the effective action $\Gamma$ for arbitrary lattice spacing 
independent of whether the transformation is local or global. 
When the transformation is global $\theta_n=\theta$, 
it is also a symmetry of the classical action $S$ 
in the continuum limit $a\to 0$ because the variation 
of the Lagrangian density induced by the transformation is 
proportional to lattice spacing. 
\begin{equation}
 \delta S = \frac{i}{2} \theta a^4 \sum_{m,n}
 \bar{\psi}_m \gamma_5 a(\slashchar{D}G-G\slashchar{D})_{m,n}\psi_n. 
\label{var}
\end{equation}
Although the variation is classically zero in the continuum limit, 
the vacuum expectation value of the variation gives the index theorem 
for arbitrary lattice spacing. 

The axial current divergence is defined as a variation 
of the classical action under the local chiral transformation 
with $\hat{\gamma}_5$. 
\begin{eqnarray}
 (\partial_\mu \hat{J}^5_{\mu})_m=\sum_{n_1,n_2}\Big[
  &&\bar{\psi}_{n_1}(\hat{\gamma}_5)_{n_1,m}\slashchar{D}_{m,n_2}\psi_{n_2}
\nonumber
\\
  +&&\bar{\psi}_{n_2}
  \slashchar{D}_{n_2,m}(\hat{\gamma}_5)_{m,n_1}\psi_{n_1}
 \Big].
\end{eqnarray}
The axial current is obtained by inverting the derivative 
\begin{eqnarray}
 (\hat{J}^5_{\mu})_n&=&
 \sum_{m}(\nabla^{-1})_{n_\mu,m_\mu}
 \prod_{\nu=1 (\nu\ne\mu)}^4 \delta_{n_\nu,m_\nu}
\nonumber
\\
&& \times
 \sum_{n_1,n_2}\Big[
  \bar{\psi}_{n_1}(\hat{\gamma}_5)_{n_1,m}(D_\mu)_{m,n_2}\psi_{n_2}
\nonumber
\\
&& \hspace{1.0cm}+\bar{\psi}_{n_2}
  (D_\mu)_{n_2,m}(\hat{\gamma}_5)_{m,n_1}\psi_{n_1}
 \Big],
\end{eqnarray}
which is gauge invariant. 
In the free theory, the currents are local in the continuum limit 
because the Leibniz rule holds in the limit. 
\footnote{In Ref. \cite{karsten} for the conventional SLAC 
derivative, axial currents are defined by introducing a derivative 
independent of the SLAC derivative, 
which is the cause of breaking of locality and 
Lorentz invariance .}
The fermion measure transforms as
\begin{equation}
 {\cal D}\psi' {\cal D}\bar{\psi}' =
 \exp\left(-2i\sum_n \theta_n {\cal A}_n \right)
 {\cal D}\psi {\cal D}\bar{\psi}. 
\end{equation}
Chiral anomaly 
\begin{equation}
 {\cal A}_n\equiv {\rm tr} (\hat{\gamma}_5)_{n,n} 
\label{anomaly}
\end{equation}
is a gauge-invariant quantity. 
As shown in Ref.~\cite{luscher},  
the index theorem holds for arbitrary lattice spacing. 
\begin{equation}
 \sum_n {\cal A}_n={\rm Index}(G). 
\end{equation}
(See also Ref.~\cite{Chiu:1998bh,Fujikawa:1999ku}.)
Since the effective action $\Gamma$ is invariant under 
transformation of the integration variables, 
the modified axial current $\hat{J}_\mu^5$ does not conserve. 
\begin{equation}
 \langle (\partial_\mu \hat{J}^5_{\mu})_n \rangle
 =-\frac{2}{a^4} {\cal A}_n. 
\end{equation}
The Ward identity for the modified chiral transformation 
holds also for the zero mode, which gives the index theorem 
and corresponds to the variation
under the global transformation (\ref{var}). 
\footnote{The explicit breaking term (\ref{var}) is 
necessary for the existence of the non-vanishing zero mode 
of the axial current divergence.}
In the continuum limit, the chiral anomaly agrees with the 
continuum result \cite{kikukawa,fujikawa2,suzuki,adams}
\begin{equation}
 \langle \partial_\mu \hat{J}_\mu^5 \rangle = \frac{1}{16\pi^2}
  \epsilon_{\mu\nu\rho\sigma}{\rm tr}(F_{\mu\nu}F_{\rho\sigma}), 
\end{equation}
where $\epsilon_{1234}=+1$ and
$F_{\mu\nu}\equiv [D_\mu^{({\rm c})},D_\nu^{({\rm c})}]$. 
The effect of chiral anomaly can be implemented to 
physical quantities via the modified anomalous axial current.

Construction of anomaly-free chiral gauge theory is easy 
if our lattice derivative is used. 
In our formulation, Weyl fermions are defined with 
the ordinary $\gamma_5$ (not $\hat{\gamma}_5$). 
As a result, the fermion measure of the path integral 
do not depend on gauge variables. 
The Weyl fermions are free from gauge anomaly beforehand. 
Consider a classical action for a Dirac fermion
\begin{equation}
 S=a^4\sum_{m,n} \bar{\psi}_m \slashchar{\hat{D}}_{m,n} \psi_n, 
\label{action2}
\end{equation}
where only the right-handed fermion is coupled to gauge. 
The left-handed fermion is redundant 
and does not contribute to the effective action. 
Once the absence of gauge anomaly is confirmed, 
the left-handed fermion can be integrated out. 
The definition of the Dirac operator $\slashchar{\hat{D}}$ 
is same as Eq. (\ref{lcd}) except that gauge is defined 
as a product of link variables
\begin{equation}
 U_{n+\hat{\mu},n}=e^{iagA_{\mu,n} P_+}, 
\end{equation}
where $A_{\mu,n}\equiv A_{\mu,n}^a T_a$ and 
$P_\pm\equiv (1\pm \gamma_5)/2$.
$\slashchar{\hat{D}}$ is no longer antihermitian. 
Under local gauge transformation, 
the classical action (\ref{action2}) is invariant. 
On the finite lattice, 
infinitesimal local gauge transformation 
\begin{equation}
 \psi_n' = (1+i\theta_n^aT_a P_+)\psi_n,\quad
 \bar{\psi}_n' = \bar{\psi}_n(1-i\theta_n^aT_a P_-),
\end{equation}
does not change the fermion measure 
\[
 {\cal D}\psi' {\cal D}\bar{\psi}'=
 \exp\left[-i\sum_n \theta_n^a {\rm tr}(T_a\gamma_5) \right]
 {\cal D}\psi {\cal D}\bar{\psi}
 ={\cal D}\psi {\cal D}\bar{\psi} 
\]
and hence the effective action. 
A single Weyl fermion can exist on the lattice 
without violating gauge symmetry. 
\footnote{The number of Weyl fermions needs to be even 
when there exists Witten's global anomaly.\cite{Witten:fp}}

\section{Summary and discussions}
\label{sumamry_and_discussions}
We have constructed a doubler-free covariant derivative 
and an anomalous Ward identity for the modified chiral 
symmetry on the lattice. 
The index theorem holds for arbitrary lattice spacing and 
the dependence of chiral anomaly on gauge fields agrees with 
the continuum result in the continuum limit. 
The zero mode of the anomalous Ward identity gives the index theorem.
In this formulation, a single Weyl fermion can exist 
on the lattice maintaining gauge symmetry.
Chiral gauge theories can be constructed nonperturbatively 
using the single anomaly-free Weyl fermion as a building block. 

The proposed lattice derivative is a simple matrix that 
gives the correct dispersion. 
We have shown that introduction of the Lanczos factor 
enables us to construct lattice derivatives with good locality. 
In this case, almost the correct dispersion is 
reproduced except for deviation at high momentum. 
It depends on fermion mass how large the truncation parameter 
$N_{\rm c}$ should be. 

The correct chiral anomaly has been obtained without violating 
locality and Lorentz invariance in the continuum limit 
in spite of the existence of long-range hopping interactions.
This is a result of a nonperturbative formulation 
based on the modified chiral transformation 
(or equivalently the modified axial current). 
To reproduce chiral anomaly, 
Wilson terms have to be introduced somewhere. 
We have introduced them only in the modified chiral transformation 
to maintain complete independence of left- and right-handed 
fermions in the classical action. 

When calculating physical quantities dependent on chiral anomaly 
such as the $\eta'$ meson mass, 
$\gamma_5$ needs to be replaced with $\hat{\gamma}_5$ 
in the concerned vertex operators to include the effect of chiral anomaly. 
It depends on the construction of the operator $G$ 
how precisely the effect of chiral anomaly is implemented 
with finite lattice spacing. 
In lattice QCD, the modification of the axial current does not affect 
physical quantities independent of chiral anomaly 
because the axial current does not couple to gauge directly.

\section*{Acknowledgments}
This research was supported in part by RIKEN.

\end{document}